\documentclass[preprint,showpacs,preprintnumbers,amsmath,amssymb]{revtex4}

\usepackage{graphicx}
\usepackage{dcolumn}
\usepackage{bm}
\usepackage{amsmath}

\begin{document}
\title{Gaslike model of social motility}
\author{A. Parravano}
\email{parravan@ula.ve}
\author{L. M. Reyes}

\affiliation{
Centro de  F\'isica Fundamental, Facultad de Ciencias,\\
Universidad de los Andes, M\'erida, Apartado Postal 26 La Hechicera, M\'erida 5251, Venezuela 
}

\date{\today}

\begin{abstract}
We propose a model to represent the motility of social elements. 
The model is completely deterministic, possesses a small number of parameters, and exhibits a series of properties that are reminiscent of the behavior of comunities in social-ecological competition; these are:
(i) similar individuals attract each other; 
(ii) individuals can form stable groups;
(iii) a group of similar individuals breaks into subgroups if 
it reaches a critical size;
(iv) interaction between groups can modify the distribution of 
the elements as a result of fusion, fission, or pursuit; 
(v) individuals can change their internal state by interaction 
with their neighbors.
The simplicity of the model and its richness of emergent behaviors, such as, for example, pursuit between groups, make it a useful toy model to explore a diversity of situations by changing the rule by which the internal state of individuals is modified by the interactions with the environment.

\end{abstract}

\pacs{05.65.+b,05.45.-a,87.23.Cc}

\maketitle

\section{Introduction}
Many living systems display collective organized displacements. Bird flocks, fish schools, animal herds, bacterial colonies, and swarms are examples of such behavior. The interactions between the traveling individuals determine the collective motion. This kind of phenomenon has been modeled by physicists 
\cite{viseck95,Toner98,Mikhailov99,helbing01,Levine01,Gregoire03,Dorsogna06,Aldana07}.
In the above models all elements in the system are identical, but Shibata and Kaneko  \cite{shibata03} considered a coupled map gas in which each element is characterized by an internal state that evolves as a result of interactions with its neighbors and additionally, the forces between element pairs depend on their internal states. In \cite{shibata03} (see also \cite{Zanette04}) the total force acting on a given element $i$ is the sum of pair forces $\vec{F}_{i,j}$ exerted by each of the elements $j$ in its vicinity, and $\vec{F}_{i,j}=-\vec{F}_{j,i}$. This kind of reciprocity is appropriate to describe many physical systems, but in social systems the reaction of an individual to its environment is usually the result of a global evaluation instead of an evaluation of the sum of neighbors' actions. In other words, the reaction of a social individual $i$ to the presence of individual $j$ does not have the same intensity as and the opposite direction to the reaction of $j$ to the presence of $i$. In particular, persecution and flight are not possible if $\vec{F}_{i,j}=-\vec{F}_{j,i}$. This kind of behavior requires a global evaluation of the environment before a decision to initiate a persecution or an escape.

This paper is organized as follows. In Sec. II we present a simple model of interacting motile elements which display pattern dynamics that evoke some basic behaviors of social comunities. The interactions of the motile element have a finite range and modify both their internal states and their positions. The functional form of the coupling of an element with its vicinity allows the formation of groups, fusion and fission of groups, and persecution and flight of groups, behavior that is observed in many species \cite{Higashi93,Bernardini96,Gat99,Gerard02,Wilson03}.
In Sec. III we analyze the properties of an isolated group as a function of its size and obtain the maximum density of elements for a compact configuration of isolated groups. In Sec. IV we show detailed simulations in a two-dimensional space for three cases of internal dynamics corresponding to (i) a homogeneous and stationary internal state, (ii) two stationary but opposite-sign internal states, and (iii) an internal dynamics with two chaotic attractors. We describe the main changes in the pattern dynamics as a function of the density of elements in the system. A summary and a brief discusion of the results and the possible applications of this model are given in Sec. V.

\section{The Social Gas model}
We propose a model of social motility that exhibits the following properties:
(i) similar individuals attract each other;
(ii) individuals can form stable groups;
(iii) a group of similar individuals breaks into subgroups if 
it reaches a critical size;
(iv) interaction between groups can modify the distribution of 
the elements as a result of fusion, fission, and pursuit;
(v) individuals can change their internal state by interaction 
with their neighbors.
We consider a system of $N$ motile elements where each element $i$ is characterized
at discrete time $t$ by a state $x_t^i$ and a position vector 
${\bf r}_t^i$ in a $d$-dimensional space.
The dynamics of the system is given by the following set of equations,

\begin{equation}
x_{t+1}^i= (1-\varepsilon)f(x_t^i)+\frac{\varepsilon}{n_t^i}\sum_{j\in \eta_t^i} f(x_t^j) \; ,
\label{est}
\end{equation}

\begin{equation}
 {\bf r}_{t+1}^i = {\bf r}_t^i + 
\gamma \,
\left(
\sum_{j\in \eta_t^i }
\frac{{\bf r}_t^j-{\bf r}_t^i}{\vert{\bf r}_t^j - {\bf r}_t^i\vert} \right)
\left(
x_{t+1}^i \sum_{j \in \eta_t^i} x_{t+1}^j
\right)
 \,.
\label{fuerza}
\end{equation}

The evolution of both the state and the vector position of element $i$ depends on its  
interactions with the elements in its neighborhood, $\eta_t^i$, such that
$\eta_t^i  =\lbrace j: \vert {\bf r}_t^j-{\bf r}_t^i \vert \leq R \rbrace$.
The parameter $\varepsilon$ expresses the coupling intensity between the states of elements, $f(x)$ is the functional form that governs the internal dynamics of each element, and $n_t^i$ is the number of neighbors of the $i$th element.

In Eq. (\ref{fuerza}), the first factor is a vector that determines the direction of motion of element $i$. Note that the influence of the neighboring elements on determining the direction of motion of $i$ depends only on their angular distribution with respect to element $i$; the greater the asymmetry of the angular distribution of neighbors, the greater the magnitude of the displacement of $i$. 
The second factor $x_{t+1}^i \sum_{j \in \eta_t^i} x_{t+1}^j$ gives the sense of motion, and it can be interpreted as the ``affinity'' of $i$ for its neighborhood; if the affinity is positive, the $i$th element will move toward the denser region of the angular distribution of neighbors, otherwise, the element will move away from this direction.
The parameter $\gamma$ expresses the coupling of an element with its neighborhood; if its value is too big, groups cannot be formed because the change $\vert {\bf r}_{t+1}^i - {\bf r}_t^i \vert$ is greater than $R$. In this way, the magnitude of the displacement depends on the position of the element relative to its vicinity and the internal states of its neighbors. If the element is isolated its position will be the same and its internal state will be determined by $f(x)$.
This model contains a simple individual strategy based on the ability of individuals to evaluate their environment and react to it. 
In most previous models the influence of neighbors on a given element can be expressed as the sum of individual contributions \cite{viseck95,Toner98,Mikhailov99,Levine01,Gregoire03,Dorsogna06,Aldana07,shibata03,Zanette04}, whereas in our model, due to the product of the two factors in Eq. (\ref{fuerza}), the effect of a neighbor on the motion of a given element depends on the states and positions of all other neighbors. This kind of holistic evaluation of the environmental conditions is a model property that allows the study of the social behavior of individuals with the ability to process the bulk of the information they acquire from their senses and react in consequence.

\section{Critical size of a group and critical density of groups}

A stable group is defined here as a set of elements that in isolation remain close during long periods of time due to the mutual interactions. 
If during evolution each element is a neighbor of the remaining elements in the group, we say that the group is a Constant Vicinity Group. On the other hand, if the neighborhood $\eta_t^i$ of element $i$ in the group is nonstationary, we say that the group is a Variable Vicinity Group. 
The diameter of a Constant Vicinity Group is always less than $R$ and it can vary in time. In contrast, a Variable Vicinity Group pulsates in a chaotic way and the maximum distance between elements frequently exceeds $R$.

To estimate the maximum size that each of the above types of group can reach, we consider a group where all the elements maintain their internal state stationary, i.e., $x_{t+1}^i=x_t^i=x$.
For a Constant Vicinity Group we estimate the maximum size $N_1$ by the condition $\vert {\bf r}_{t+1}^i - {\bf r}_t^i \vert \leq R$; that is, an element $i$ in the periphery of the group remains a neighbor of the rest of elements in the group in the next time step. Assuming a uniform distribution of elements on a disk of diameter $R$, Eq.~(\ref{fuerza}) yields
\begin{equation} 
N_1 \quad \approx  \quad \sqrt{\frac{\pi}{2} \frac{R}{\gamma x^2}} -1\; .
\label{Ncri1}
\end{equation}

In the case of a Variable Vicinity Group, the maximum size $N_2$ can only be roughly estimated
due to the non-uniform distribution of the elements. 
Assume that at time $t$ the group is very asymmetric and the maximum distance between elements is less than $R$. Then, at time $t+1$ this group splits into two or more noninteracting groups when the condition 
$\gamma x^2 ( N - 1 )^2 > 2R$ is satisfied. Therefore, the maximum size of a Variable Vicinity Group is given by
\begin{equation} 
N_2 \quad \approx  \quad \sqrt{\frac{2R}{\gamma x^2}} -1\; .
\label{Ncri2}
\end{equation}

\begin{figure*}
\hbox{\includegraphics[scale=0.6,angle=90]{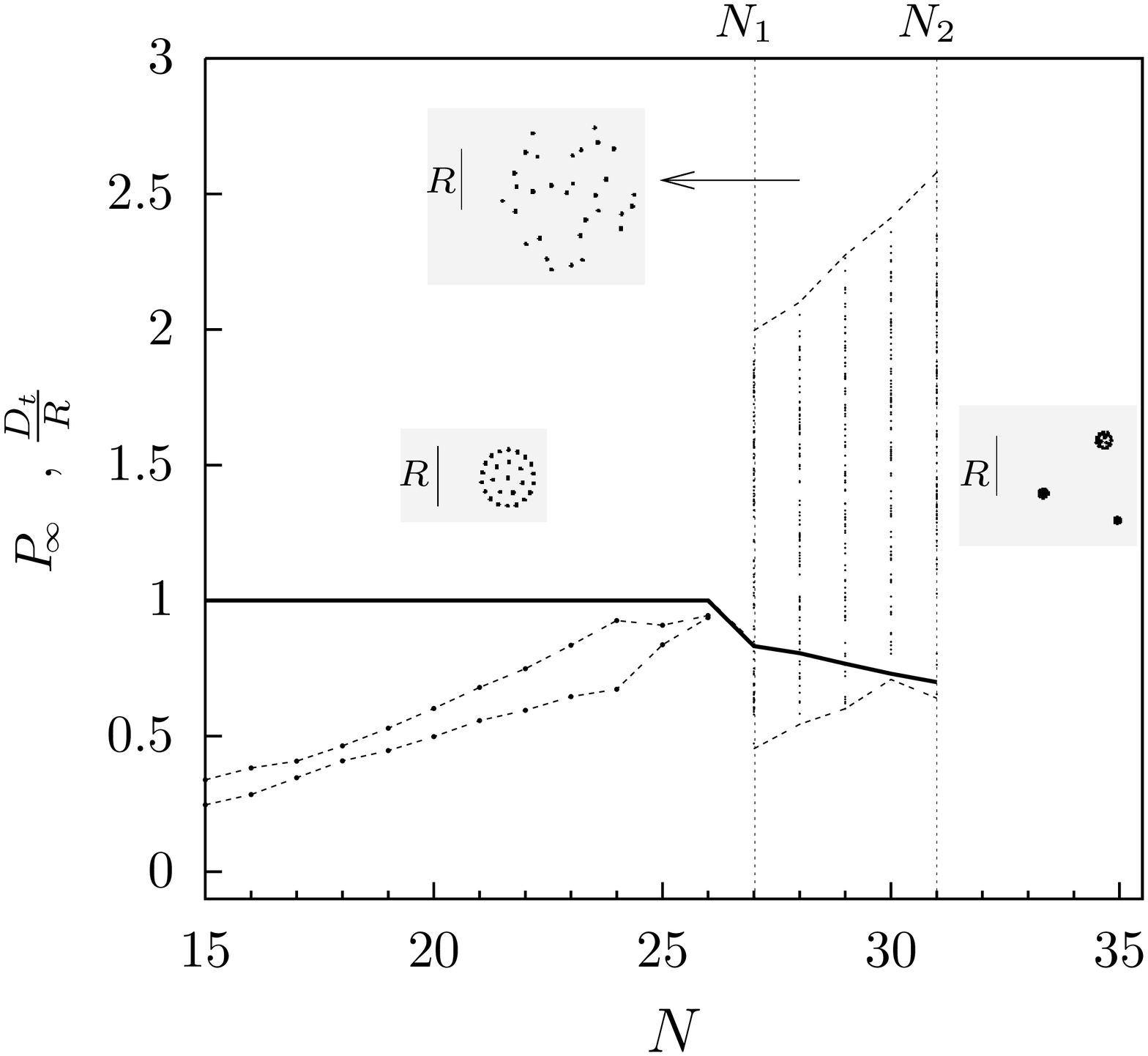}}
\caption{\label{ncri} 
$P_\infty$ (continuous curve) and $D_t/R$ (dots) as functions of the group size $N$ for parameter values $R=5$, $\gamma=0.01$, and $x=1$. The critical sizes $N_1$ and $N_2$ are indicated. The insets show the typical configurations for $N<N_1$, $N_1<N<N_2$, and $N>N_2$.}
\end{figure*}

In order to characterize the transition from a Constant to a Variable Vicinity Group,  we define the quantity
\begin{equation}
P(\tau)=\frac{1}{\tau}\sum_{t=1}^{\tau} \frac{1}{N}\sum_{i=1}^{N} \frac{n_t^i}{N-1}  \; ,
\label{Pg}
\end{equation}
which describes the average fraction of elements that belong to a neighborhood, where the time average is calculated after discarding a number of transients. 
Figure \ref{ncri} shows the asymptotic quantity $P_\infty$ as a function of $N$, with fixed 
values of parameters $R=5,\gamma=0.01$, and $x=1$. 
For $N<N_1=27$, the system forms a Constant Vicinity Group for which $P_\infty=1$. At $N=N_1$ the system experiences a transition to a variable vicinity group for which $P_\infty < 1$. At $N=N_2=31$ the system undergoes a structural change, resulting in the splitting into two or more groups. The values of $N$ at which the above transitions occur agree with the theoretical values of $N_1$ and $N_2$ given by Eqs. \ref{Ncri1} and \ref{Ncri2}, as indicated in Fig.~\ref{ncri}.

The temporal behavior of the system can be characterized by the  
distance between the two furthest separated elements in the group at time $t$, denoted by $D_t$.
Figure \ref{ncri} also shows the bifurcation diagram of $D_t$ as a function of $N$.
For each value of $N$, $5000$ iterations were discarded as transients and the next 100 values of $D_t$ were plotted in Fig.\ref{ncri}. The size of the system was increased by adding one element at a random position within a distance $R$ of the center of the group of size $N$. 
We observe that $D_t$ displays a period two for $N < N_1$, as a result of the pulsatory behavior of the size of the constant vicinity group. For $N_1< N < N_2$, $D_t$ shows a chaotic behavior corresponding to a variable vicinity group. At $N = N_2+1$, the group becomes unstable and in a few iterations it splits into two or more stable groups. The insets in Fig.~\ref{ncri} show typical configurations of the system for the values of parameter indicated. 
 
Now consider the situation when $N$ elements are distributed on an area of size $L \times L$. When the density $\rho=N/L^2$ is below a critical value, the elements can self-organize into different configurations of noninteracting stable groups. In particular, configurations of identical stable groups of size $N_1$ can support the largest number of elements in the system.
The maximum number of noninteracting groups of size $N_1$ that can exist in a compact arrangement is
$N_g=(L/R)^2/(2\sqrt{3})$, so the mean density of elements in the system is
\begin{equation}
\rho_1 \; = \frac{N_1 N_g}{L^2} =\; \frac{\sqrt{\frac{\pi R}{2 \gamma x^2}}-1}{2\sqrt{3}R^2} \; ,
\end{equation}

In spite of the fact that $N_2$ determines the critical number above which an isolated group splits, when the interactions between groups are considered the relevant quantity to estimate the maximum density of elements that the system can suport in noninteracting groups is $\rho_1$ 
 (i.e. $\rho_2=2 (\sqrt{\frac{2 R}{\gamma x^2}}-1)/(9\sqrt{3}R^2) \sim \rho_1/2$).
However, a stable configuration with all groups of the same size is reached only with very particular initial conditions. For most initial conditions the groups in the final stable configurations have a variety of sizes. Consequently, in general the configurations become unstable at a critical density below $\rho_1$. 
From now on, we express the surface densities of elements in the simulation area, $\rho=N/L^2$, in terms of the normalized density $\tilde{\rho}=\rho / \rho_1$

\section{Social Gas behavior}
To illustrate the main properties of the social gas model we analyze its behavior for three cases of the internal dynamics $f(x)$ in a two-dimensional system with periodic boundary conditions. 
We focus on the dependence of the pattern dynamics on the particle density $\rho$ and show results for a particular set of model parameters, namely, $R=10$ and $\gamma=0.01$.
 
The collective behavior of the system is characterized here by  
the average number changes of neighbors per element during a number of iterations $\tau$, 
\begin{equation}
S (t,\tau)=
\frac{1}{\tau} \sum_{t}^{t+\tau} \frac{1}{N} \sum_{i=1}^N 
\left[ \Delta n_t^i (+) + \Delta n_t^i (-)\right]  \; ,
\label{ecpromveci}
\end{equation}
where $\Delta n_t^i (+)$ ($\Delta n_t^i (-)$) is the number of elements that enter (leave) the vicinity of element $i$ in the time step from $t$ to $t+1$.
In the following, three cases of internal dynamics $f(x)$ are considered: (i) the state of all elements is stationary and homogeneous; (ii) the state of elements is stationary with one half of them having the state $x_{n}^i=1$ and the other half $x_t^i=-1$; and (iii) the states of elements evolve following Eq. (\ref{est}) with an internal dynamics described by the Miller-Huse map \cite{lemaitre}:
 \begin{displaymath}
f(x)= \left \lbrace \begin{array}{rrrrrrrrr}
-2a/3 - ax, & & if & x & \in & (-1,-1/3)\\ ax, & & if & x &\in& (-1/3,1/3)\\ 2a/3 - ax, & & if & x & \in & (1/3,1) \end{array} \right . \; .
\label{mapa}
\end{displaymath}

\begin{figure*}
\hbox{\includegraphics[scale=1.0,angle=0]{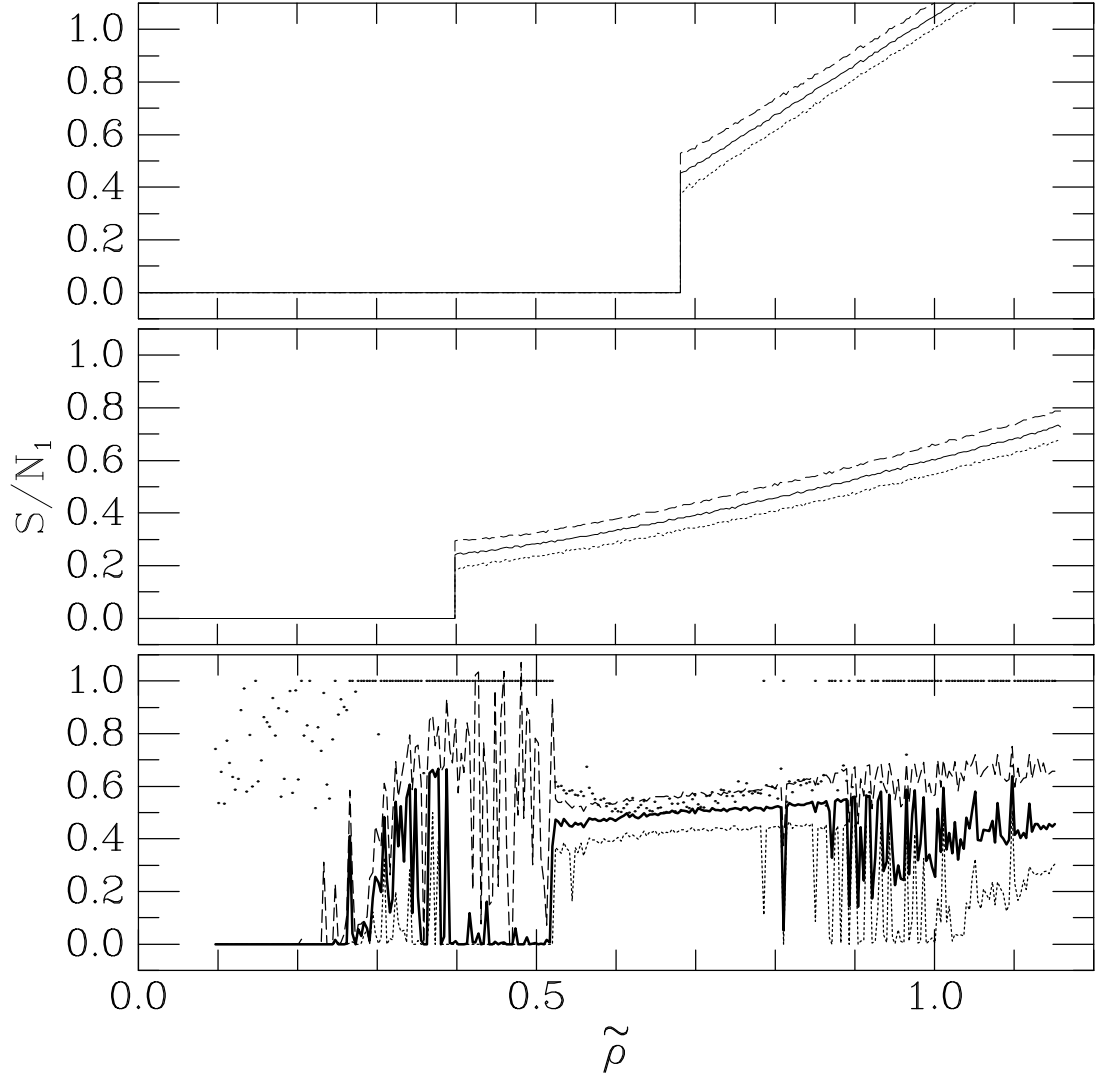}}
\caption{\label{perdesv} 
Average number changes of neighbors per element as function of density when $R=10$, $\gamma=0.01$, and $L=50$. The continuous curve corresponds to the median value $S_{100,med}$, whereas the dotted and dashed curves correspond, respectively, to the percentiles $S_{100,10\%}$ and $S_{100,90\%}$; see text. Starting from random initial conditions, these quantities are calculated in the period $10^4\, <\, t\, <\,6\times 10^4$ and are normalized to the critical size $N_1$ ($=39$ for these parameter values).
Top panel: the state of all elements is stationary and homogeneous, i.e., $f(x_{t}^i)=1$. 
Middel panel: the state of elements is stationary with one half of them having the state $x_{n}^i=1$ and the other half $x_t^i=-1$.
Bottom panel: the internal dynamics is given by the Miller-Huse map with $a=1.9$ and $\varepsilon=0.3$;
the dots correspond to the fraction $f_{band}$ of elements with the majority sign.
}
\end{figure*}

\begin{figure*}
\hbox{\includegraphics[scale=1.4,angle=0]{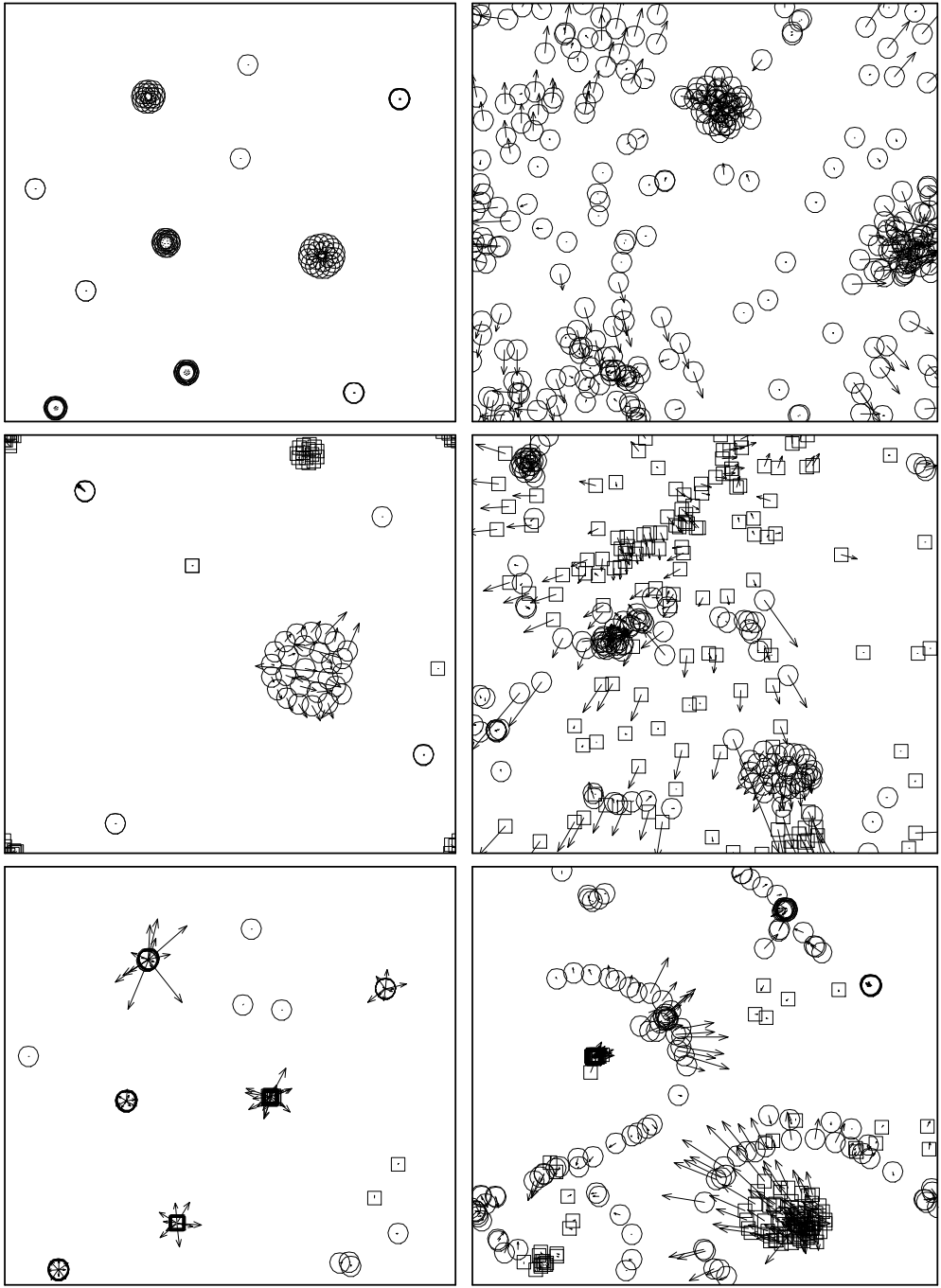}}
\caption{\label{slides} Typical configurations at low density ($\tilde{\rho}=0.3$, left hand panels) and intermediate density ($\tilde{\rho}=1$, right hand-panels) at a particular iteration $t_s$. Top, middle and bottom panels correspond to the three cases in Fig. \ref{perdesv}. 
The positions of particles at time $t_s$ are represented with circles for positive states and squares for negative states. The arrows represent the displacement vectors between the positions at iterations $t_s-1$ and $t_s+1$.}
\end{figure*}

For these three cases Fig. \ref{perdesv} summarizes the behavior of the system 
by means of the quantity $S_{100}(t)\equiv S(t,\tau=100)$ calculated each $100$ iterations of $t$. The median $S_{100,med}$ of $S_{100}$ and the value $S_{100,10\%}$ ($S_{100,90\%}$) below which $S_{100}$ fall $10\%$ ($90\%$) of times are shown as functions of $\tilde{\rho}$. Figure \ref{slides} shows snapshots of the system state at particular values of $\tilde{\rho}$. The main results are as follows.

(i) The case where the state of all elements is stationary and homogeneous, i.e. $f(x_{t}^i)=1$, is shown in the top panel of Fig. \ref{perdesv}. Note that $S_{100,med}$ exhibits a discontinuous transition at $\tilde{\rho}_{c}\approx 0.7$ from a configuration of noninteracting stable groups with constant vicinity (as shown in the left top panel of Fig. \ref{slides} and characterized by $S_{100}=0$) to a configuration where elements tend to form groups with variable vicinity that become unstable due to sporadic interactions with neighbor groups (as shown in the right top panel of Fig. \ref{slides} and for which $S_{100}>0$). This first-order transition occurs because, as soon as one of the Constant Vicinity Groups exceeds the critical size $N_1$, it becomes a Variable Vicinity Group and almost duplicates its radius. In this situation, the groups in the system can no longer remain isolated and they exchange elements. In general, the transition is observed at $\tilde{\rho}_{c} < 1$. Stable configurations for $\tilde{\rho} \lesssim 1$ can be reached only for very specific initial conditions that yield to a set of isolated stable groups all having sizes $\lesssim N_1$. For $\tilde{\rho} < \tilde{\rho}_{c}$ the system evolves from the disordered initial random conditions to a stable configuration like the one shown in the left top panel of Fig. \ref{slides}. For $\tilde{\rho} < \tilde{\rho}_{c}$ most initial conditions yield stable configurations of Constant Vicinity Groups (i.e., $S=0$); however, cases with one or more isolated Variable Vicinity Groups can occur (i.e., $S\gtrsim 0$). The number of iterations necessary to reach a stable configuration quickly increases as $\tilde{\rho}$ approaches $\tilde{\rho}_{c}$. Reference \cite{Movies} shows 2D movies of the evolution of the system, corresponding to the cases shown in the top panels of Fig. \ref{slides}.

(ii) The case where the state of elements is stationary with one-half of them having the state $x_{n}^i=1$ and the other half $x_t^i=-1$ displays a behavior that is reminiscent of two communities that chase each other.
Initially the elements are distributed at random in the simulation area; they soon tend to segregate into a number of  groups, each group having members of the same sign. If the density is low enough, isolated stable groups soon form, but as the density increases the interaction among groups makes it difficult to reach a stable configuration. The existence of states of opposite sign allows for attractive or repulsive interactions between groups. These interactions cause the larger groups to pursue and disperse the smaller groups of contrary sign in frontlike configurations. At the same time, attraction among elements of the same sign tends to increase the size of existing groups and to form new ones; however, if a group reaches a size $N_2$ the group splits. Groups of opposite signs having similar size attract each other; as a result both groups end up dispersing.
 As shown in the middle panel of Fig. \ref{perdesv}, $S_{100,med}$ exhibits a discontinuous transition at
 $\tilde{\rho}_{c}\approx 0.4$. As stated above, for $\tilde{\rho}<\tilde{\rho}_{c}$, each community forms isolated
 Constant Vicinity Groups consisting of elements with the same state (as shown in the middle left panel of Fig.
 \ref{slides}). For $\tilde{\rho}>\tilde{\rho}_{c}$ a chasing behavior emerges in the system (as shown in the
 middle right panel of Fig. \ref{slides}). 
As $\tilde{\rho}$ increases the system displays a large degree of disorder, sustained by frequent interactions between elements of contrary sign that inhibit the persistence of well-defined groups. Movies of the evolution of the system, corresponding to the cases shown in the middle panels of Fig. \ref{slides} are also available at \cite{Movies}.

\begin{figure*}
\hbox{\includegraphics[scale=1.0,angle=0]{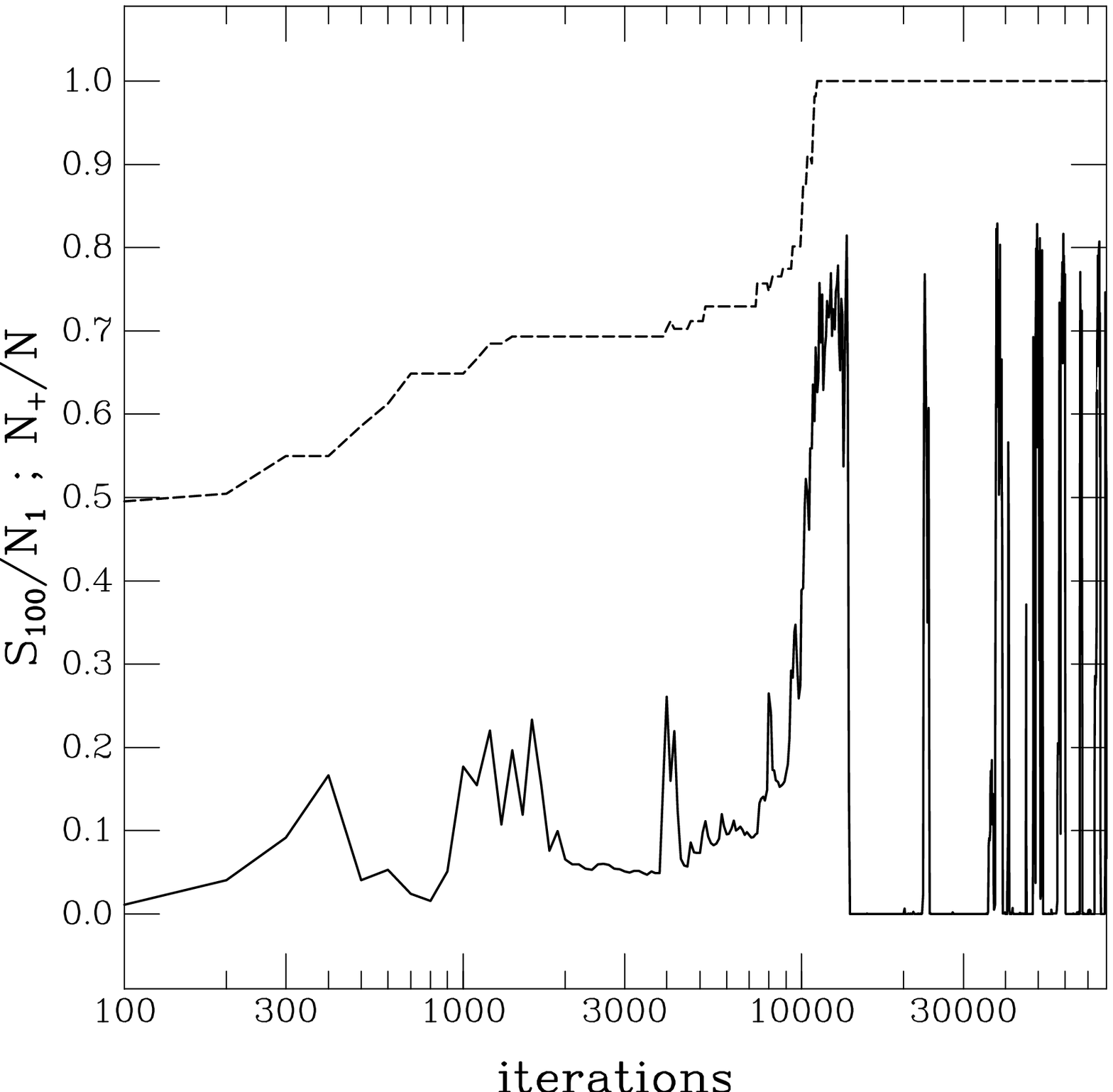}}
\caption{\label{evo} Evolution of $S_{100}$ normalized to the critical size $N_1$ (continuous curve) and the evolution of the fraction of elements with positive states (dashed curve). The simulation corresponds to the case when $\tilde{\rho}=0.4$ and the internal dynamics is given by the Miller-Huse map with parameter values as in the bottom panel of Fig. \ref{perdesv}.}
\end{figure*}

(iii) In the last case the states of elements evolve following Eq. (\ref{est}) with the internal dynamics given by the Miller-Huse map.
When $a \in (1,2)$, this map has two symmetric chaotic band attractors, one with values $x_t^i>0$ and the other with $x_t^i<0$, separated by a finite gap centered at $x=0$. For parameter values $a=1.9$ and $\varepsilon=0.3$, the gap is small and interactions are sporadically strong enough to produce element sign switching. In addition to the chasing phenomenon observed in case (ii), now the fleeing elements may change their sign; a scenario reminiscent of a process of conversion or assimilation of individuals by larger comunities. In contrast to cases (i) and (ii), in case (iii) the isolated groups display randomlike motions (i.e., the chaotic evolution of the state of the elements in a group result in a fluctuating motion of the geometrical center of the group).
For these parameter values the botton panel of Fig. \ref{perdesv} shows $S_{100,med}$, $S_{100,10\%}$ and $S_{100,90\%}$ as functions of $\tilde{\rho}$, as well as the fraction $f_{band}$ of elements that at the final iteration ($t=6 \times 10^4$) are in the more populated chaotic band, the positive or the negative. Even when the average value of the state of elements in a positive group is less than 1, in this case $\rho$ is normalized to the same value $\rho_1$ used in cases (i) and (ii). In contrast to cases (i) and (ii), the system never reaches long-term stationary configurations because, even at low densities, the isolated groups move and eventually interact with other groups, resulting in an exchanges of neighbors (i.e., $S>0$), and sometimes the internal state of one or more elements change their sign. As shown in the botton panel of Fig. \ref{perdesv} there is not a clear transition as in cases (i) and (ii). Note the change of behavior when all the elements end up in one band, i.e., $f_{band}=1$.
In these simulations $f_{band}\approx 0.5$ for $0.55 \lesssim  \tilde{\rho} \lesssim 0.85$. In this range
of $\tilde{\rho}$ there is exchange of elements between the two chaotic bands, but an important majority is rarely established, and consequently the behavior is similar to case (ii). 
For $\tilde{\rho} \gtrsim 1$ transient stable configurations (i.e., $S_{100}=0$) are not longer attained. The botton panels of Fig. \ref{slides} show snapshots of typical configurations for low and high densities (see also \cite{Movies}). Figure \ref{evo} shows the evolution of $S_{100}$ for $\tilde{\rho}=0.4$ as well as the fraction of elements with positive states. Note that at the beginning the number of positive elements is in a slight minority, but they soon become the majority, and finally all elements end positive. As long as $f_{band}(t)<1$, the evolution of $S_{100}$ shows quiet periods ($S\sim 0$) followed by active periods that are usually associated with sign changes. As soon as $f_{band}(t)=1$,
the active periods reach higher values of $S_{100}$ and the quiet periods are characterized by $S_{100}=0$. 
 This is why in the botton panel of Fig. \ref{perdesv} $S_{100,10\%} \sim 0$ when $f_{band}=1$ and
$\tilde{\rho} < 1$. For $\tilde{\rho} \gtrsim 1$ quiet periods are absent ($S_{100,10\%} > 0$) because 
quasistable configurations are no longer possible.

\section{Summary and Conclusions}
We have analyzed the evolution of the patterns displayed by a coupled map gas in which the state and position of elements vary as a result of the interactions with their neighbors. The motion of elements is governed by a rule inspired by the fact that the reaction of a social individual to its environment is the result of a global evaluation instead of an evaluation of the sum of the neighbors' actions. The proposed model is completely deterministic, possesses a small number of parameters, and exhibits a series of properties that are reminiscent of the behavior of comunities in competition (i.e., fission, fusion, and pursuit of groups of elements). 
The pattern dynamics depend on the model parameters $\gamma$ and $R$ and on the density of elements $\rho=N/L^2$ in the system area. We have analyzed the behavior of the system for three cases of the internal dynamics corresponding to (i) a homogeneous and stationary internal state, (ii) two stationary but opposite internal states and (iii) an internal dynamics with two chaotic attractors of opposite sign. In cases (i) and (ii) there is a transition at a critical value of the density from a stable configuration (quiet mode) to a pattern of interacting groups (conflict mode); in case (ii), pursuit and flight of groups is the dominant feature in the conflict mode. In case (iii) isolated groups display ramdonlike displacements of their geometrical center and consequently even at low densities the system displays quiet periods separated by periods of conflict. The duration of the conflict periods increases with $\rho$; and for high enough densities quiet periods are absent. Additionally, in case (iii) the sign of the elements may change and in many cases all the elements end up trapped in one of the chaotic bands. 

The results in this paper indicate that this kind of simple and deterministic model might be used to study basic properties of the collective social behavior. 
It must be noted that this study is by no means complete. Qualitatively different behaviors are expected for other sets of parameters and internal dynamics $f(x)$. For example, for negative values of the parameter $\gamma$ an isolated pair of elements repel (attract) if the elements have the same (opposite) sign; if all elements have the same sign stable groups cannot form, but when both signs are present an interesting behavior arise in which quasistable inhomogeneous groups form. If a periodic or quasiperiodic internal dynamic is adopted, the formation of synchronized groups is expected for appropriate parameter values. If the internal dynamics is multidimensional, as in the Axelrod model of cultural dissemination \cite{Axelrod97,Gonzalez-Avella05,kuperman06}, the interaction among elements can be designed to include more sophisticated relations.

\begin{acknowledgments}
We thank Mario Cosenza for useful comments that helped to improve the manuscript.
This work was supported by Consejo de Desarrollo Cient\'ifico,
Human\'istico y Tecnol\'ogico of the Universidad de Los Andes,
M\'erida,  under grants No. C-1275-04-05-B  and C-1271-04-05-A. 
\end{acknowledgments}

\eject

\begin{thebibliography}{99}
\bibitem{viseck95}T. Vicsek, A. Czir\'ok, E. Ben-Jacob, I. Cohen, and O. Shochet. Phys. Rev. Lett. 75, 1226 (1995)
\bibitem{Toner98}J. Toner and Y. Tu, Phys. Rev. E 58, 4828 (1998).
\bibitem{Mikhailov99} A.S. Mikhailov and D.H. Zanette. Phys. Rev. E 60, 4571 (1999).
\bibitem{helbing01}  D. Helbing, Rev. Mod. Phys. \textbf{73}, 1068 (2001).
\bibitem{Levine01} H. Levine, W.J. Rappel, and I. Cohen, Phys. Rev. E 63, 017101 (2000).
\bibitem{Gregoire03} G. Gregoire, H. Chate, and Y. Tu. Physica D 181, 157 (2003).
\bibitem{Dorsogna06}M. R. DOrsogna, Y. L. Chuang, A. L. Bertozzi, and L. S. Chayes, Phys. Rev. Lett. 96, 104302 (2006).
\bibitem{Aldana07} M. Aldana, V. Dossetti, C. Huepe, V. M. Kenkre, and H. Larralde, Phys. Rev. Lett. 98, 095702 (2007).
\bibitem{shibata03} T. Shibata and K. Kaneko, Physica D \textbf{181}, 197 (2003).
\bibitem{Zanette04} D. H. Zanette and A. S. Mikhailov, Physica D 194, 203 (2004).
\bibitem{Higashi93} M. Higashi and N. Yamamura, Am. Nat. 142, 553 (1993).
\bibitem{Bernardini96} W. Bernardini, J. Anthropol. Archaeol. 15, 372 (1996).
\bibitem{Gat99} A. Gat, J. Anthropo. Res. 55, 563 (1999).
\bibitem{Gerard02} J.F. Gerard, E.Bideau, M.L. Maublanc, P. Loisel, and C. Marchal, Biol. Bull 202, 275 (2002).
\bibitem{Wilson03}M.L. Wilson and R.W. Wrangham, Annu. Rev. Anthropol. 32, 363 (2003).
\bibitem{lemaitre} A. Lemaitre and H. Chat\'e, Phys. Rev. Lett. \textbf{80}, 5528 (1998).
\bibitem{Axelrod97} R. Axelrod, \textit{The complexity of coperation. Agent-based models of competition and collaboration} (Princeton University Press, Princeton, NY, 1997).
\bibitem{Gonzalez-Avella05} J. C. Gonz\'alez-Avella, M. G. Cosenza, and K. Tucci, Phys. Rev. E 72, 065102(R) (2005).
\bibitem{kuperman06} M.N. Kuperman, Phys. Rev. E 73, 046139 (2006).
\bibitem{Movies}See http://webdelprofessor.ula.ve/ciencias/parravan/ movi-irho0p3.ps and movi-i-rho1p0.ps for the top panels of Fig. 3; movi-ii-rho0p3.ps and movi-ii-rho1p0.ps for the middle panels of Fig. 3; and movi-iii-rho0p3.ps and movi-iii-rho1p0.ps for the bottom panels of Fig. 3.
\end{thebibliography}
\end{document}